\documentclass[12 pt]{article}

\usepackage{courier}

\usepackage{amsmath}
\usepackage{caption}
\usepackage{graphicx}
\usepackage{relsize}
\usepackage{xcolor}
\usepackage{appendix}
\usepackage{multirow}
\usepackage{multicol}
\usepackage[normalem]{ulem}
\usepackage{sectsty}
\sectionfont{\centering}
\subsectionfont{\centering}
\usepackage[english]{babel}
\usepackage[utf8]{inputenc}
\usepackage{setspace}
\usepackage{bm}
\usepackage{hyperref}
\usepackage[T1]{fontenc}

\usepackage[top=2cm, bottom=2cm, right=2.0cm, left=2.0cm]{geometry}
\usepackage{authblk}

\title{Study of ion induced Inner Shell Ionization cross section through electron capture mechanism}
\author[1]{Sumana Ghosh}
\author[1]{Debasis Mitra}
\author[2]{Soumya Chatterjee}
\affil[1]{\textit{Dept. of Physics, University of Kalyani, Kalyani, Nadia-741235, India.}}
\affil[2]{\textit{Dept. of Physics, Brainware University, Barasat, Kolkata-700125, India.}}
\date{}
\begin{document}


\maketitle
\doublespacing
\begin{abstract}
Electron Capture (EC) cross-section from K, L and M shells of the target atoms to the vacant K, L and M shells of the projectile ions have been calculated by deriving the accurate momentum transfer to the captured electrons for different charge states. Several correction factors like polarization correction, relativistic effects (R) of the target wave function, Coulomb-deflection factor (C) due to the effect of the repulsion between the projectile and the target nucleus, correction for projectile energy loss have been introduced. 
The mean charge state of the projectiles inside the target material has been estimated using suitable empirical models and the fractional charge state distribution has been calculated considering Lorentz distribution. Fractional distribution of charge state of the projectile ions is used to obtain the charge state contributions of the electron capture cross-sections. The effect of Simultaneous Multiple Ionization (SMI) has been considered in the theory of Direct Coulomb Ionization (DCI). The  theoretically obtained total cross-sections have been compared with the experimental findings obtained from various literature.  The computation scheme has been depicted through sample calculations of ionization cross-sections through electron capture mechanism.
\end{abstract}
\section{Introduction}
Over the past few decades, various theoretical studies and experiments have confirmed that apart from direct Coulomb Ionization, inner-shell ionization of the target atoms can take place as a result of their electrons being captured by the colliding projectile ions during ion-atom collisions.  The projectile charge state dependence of x-ray production cross-section (particularly for heavy projectile ions) has been studied in various experiments \cite{macdonald1972dependence,mowat1972projectile,brandt1973dynamic}  and significant differences are observed from the predicted dependence on $z_1^2$ ($z_1$ is nuclear charge of the projectile) in Coulomb ionisation formula. The most successful theory of Direct Coulomb Ionization (DCI), which was developed from PWBA by incorporation of Binding and Polarisation correction factor, Coulomb deflection effect, Energy loss Effect and Relativistic correction factor, known as ECPSSR theory (for low energy ECUSAR) \cite{basbas1971projectile,brandt1966binding,brandt1966characteristic} can explain the experimental inner-shell ionization data for light projectile ions like proton, alpha particle etc. However, for heavy ion induced inner-shell ionization, the theoretical predictions give significantly low value compared to the 
 \cite{msimanga2016k, gluchshenko2016k, mitra2010lower} experimental data.  Consideration of the change in the fluorescence yield and Coster-Kronig transition rates due to the creation of multiple vacancies in target atom during the heavy ion induced collision \cite{richard1969observed,burch1970x,knudson1971aluminum,richard1972direct,burch1972effect} can not bridge the gap between the experimental results to the direct Coulomb ionisation theory. So, the inclusion of the inner-shell vacancy creation due to the  target electron capture by the projectile ions might be a possible solution to resolve the discrepancy between the theory and experimental data, particularly in case of comparable binding energy between target active shell and the projectile vacant shell where the electron is being transferred. Halpern and Law \cite{halpern1973k}, have studied Argon K-shell ionisation induced by several bare projectile ions (e.g. C, F etc.) and shown that the consideration of the contribution of ionisation of target atom due to the electron transfer to projectile ions obtained from OBK formalism with appropriate scaling factors make better agreement with the experimental results. However, no physical justification of the scaling factors that have been multiplied to the OBK capture cross-sections was given in the paper. Nikolaev  \cite{nikolaev1967calculation} further modified the OBK formalism using non-relativistic screened Hydrogenic wave function for the bare projectile ions named as OBKN electron capture cross-section and no scaling is required in this formulation.
Lapicki et al. \cite{lapicki1977electron,lapicki1980electron} have extended the above mentioned OBKN formula for medium to low projectile velocity in case of electron capture from target K-shell. For low projectile velocities, Coulomb deflection of projectile ions by target nucleus and the change in binding energy of the target active electron has been introduced.The large binding energies of electrons in the K and L shells of medium and heavy z targets ensure the high velocity of target electrons, necessitating the relativistic aspect of the electronic wave-function, which have also been taken into account by Lapicki et al. \cite{lapicki1980electron}. However,  the formalism was developed for fully stripped projectile ions and only for the K-shell of the target atom. The  above prescription shows good agreement with experimental results so far as the gaseous target is concerned. But, in case of solid target the picture is different and more complicated. While traversing through the solid matter, projectile ions acquire and loose electrons due to the interaction of the free electrons inside target material. As a result, a distribution of charge states of the projectile ions is manifested inside the target. So, the simplistic form of momentum transfer to the exchanged electron as prescribed by Lapicki et al. can not be considered for collision with solid targets. During the inelastic collision, projectile ions loose energy which affects the ionization process through electron capture similar to the direct Coulomb ionisation \cite{brandt1981energy} and this phenomena has not been included in the Lapicki's formalism of electron capture. The polarization of the target atom due to the positively charged projectile ions also distorts the quantum mechanical wave functions causing a change in the electron capture cross-section which has also not been considered. We have addressed the above-mentioned correction factors and redefined the process of momentum transfer to the captured electron to get a more precise approach of the target electron capture formalism. The formulation has not only been extended for target K-shell but also been derived for sub-shell resolved L and M shell. Although the inclusion of these correction factors are based on the approach of Lapicki et al.\cite{lapicki1980electron}, it should not be viewed as a mere modification of the formula. In the following sections we have shown how we have incorporated the desired correction factors and step by step developed the electron capture cross-section from target K-shell and sub-shell resolved L and M shells. 
\section{Analytical calculation of Electron Capture (EC) cross-section}
The cross-section of electron capture from the target inner-shells has been derived by Nikolaev \cite{nikolaev1967calculation} using screened Hydrogenic wave function in the OBK formalism known as OBKN cross-section. 
Lapicki et al. \cite{lapicki1980electron} have represented the above formalism for K-shell in terms of reduced binding energy $\theta_t$ which has been modified for all target shells (i.e. K, L and M) in case of fully stripped ions and  written as,
\begin{equation}
\sigma^{OBKN} = \frac{2^8 a_0^2 N}{5}\pi(\frac{n_1}{v})^2(\frac{v_{1}}{v_{2}})^5\xi_t^{10}(\theta_t)\frac{\phi_4(\zeta)}{(1+\zeta)^3}\label{OBKN}
\end{equation}
where,
\begin{equation}
\theta_t=\frac{n_2^2}{z_{2s}^2}\frac{U}{13.6} \label{theta}
\end{equation}
\begin{equation}
\zeta=(1-\theta_t)\xi_t^2(\theta_t)\label{zeta}
\end{equation}
\begin{equation}
    q_t(\theta_t)=\frac{1}{2}(v+\frac{|v_2^2\theta_t-v_1^2|}{v})\label{q(theta)}
\end{equation}
\begin{equation}
\xi_t(\theta_t)=\frac{v_2}{[v_1^2+q_t^2(\theta_t)]^{1/2} }\label{xi(theta)}
\end{equation}
\begin{equation}
\phi_4(\zeta)=\frac{1}{1+0.3 \zeta}\label{phi_theta}
\end{equation}
here, $U$ is the binding energy and $N$ is the number of electrons exist in the target active shell (or sub-shell for sub-shell resolved cases), $v$ is the velocity of the incident projectile ion, $n_1$ is the principal quantum number of projectile where the electron is being captured. $v_1$ and $v_2$ are the velocities of electron in atomic units for projectile and target active shell respectively. $v_2$ is obtained as $v_2=z_{2s}/n_2$ where, $z_{2s}$ is screened nuclear charge and $n_2$ is the principle quantum number of target active shell from which electron is shifted. 
The shielding effect of the existing electrons on the nucleus of the target atom has been taken care of by modifying the nuclear charge using screening constants as given by Slater
\cite{slater1930atomic} as $z_{2K}=z_2-0.3$, $z_{2L_1}=z_{2L_2}=z_{2L_3}=z_2-4.15$, $z_{2M_1}=z_{2M_2}=z_{2M_3}=z_2-11.25$ and $z_{2M_4}=z_{2M_5}=z_2-21.15$. $q_t(\theta_t)$ is the minimum momentum transfer for electron capture process. 
 Unlike Direct Coulomb Ionisation the electron velocity in the projectile active shell is necessary for obtaining the proper minimum momentum transfer to the captured electron. In case of bare projectile ions Lapicki et al. \cite{lapicki1980electron} have considered $v_1$ to be $z_1/n_1$. The minimum momentum transfer $q_t(\theta_t)$ for different projectile charge states has been discussed in the subsection 2.1.
\subsection{Minimum Momentum transfer of the exchanged electron due to different projectile charge states}
The existence of the electrons in the projectile ions of different charge states decreases the binding energy of the vacant shells due to the shielding effect compared to the fully stripped projectile ions. The Slater's screening constants can not provide the effective projectile nuclear charge acting on the captured electron as the projectile ion is not completely filled. So we have used a different technique where we have considered the ionisation potential of ions in its ground state as prescribed by Agmon \cite{agmon1988ionization} after the electron has been captured (which is nothing but the binding energy of the captured electron) in the inner most vacant shell of the projectile ion. And from this binding energy we can easily calculate the velocity of the said electron. The ionisation potential is calculated in equation (1) of \cite{agmon1988ionization}. 
So, the velocity of the captured electron in the inner most vacant shell of the projectile ion can be calculated from the given equation and also we can easily calculate the velocity of the electrons in higher shell of the projectile ions through simple scaling. 

For the projectile approaching with low velocities it is necessary to modify the OBKN cross-section formula for capture as in the case of Direct Coulomb Ionization ECPSSR theory \cite{liu1996isics}.  In the next sub-sections we have discussed the modification of OBKN formula by introducing the correction factors for slow collision domain.
\subsection{Binding and polarisation correction}
The impact of the slow projectile ions on the electrons in the target shell has been derived by perturbed-stationary-state theory \cite{basbas1973perturbed}. Lapicki and McDaniel \cite{lapicki1980electron} have introduced the effect of the change in binding energy of the target shell electron due to the presences of projectile ion in the capture cross-section formula for K-shell. We would like to extend this formulation of binding energy correction for sub-shells of both L and M shells. We also considered the correction due to the polarisation effect in the capture cross-section not only for K-shell but also for sub-shell resolved L and M shells of the target atom.   

The binding effect is incorporated in the capture cross-section formula by introducing the binding correction term $\epsilon_t^B$ as given by Brandt and Lapicki in their theory of Direct Coulomb ionisation in equation (17) for K and L shells 
where, the term $g_t$ in the binding correction factor for is given in equation (19) of \cite{brandt1979shell}.
Here, the factor $\xi$ in $g_K, g_{L_1}, g_{L_2L_3}$ is given in equation (\ref{xi(theta)}) of this paper.

For large impact parameter the trajectory of the projectile ion is far from the active shell of the target atom causing perturbation in the state which leads to the polarization effect. The correction factor of this polarisation effect has been derived in ECPSSR theory and we have considered it in the capture cross-section formula in a similar way. The polarisation correction factor for K and L-shell is equation (16) as given in \cite{brandt1979shell} 
The adjustable parameter $c_s$ in the I is introduced by Merzbacher et al.  \cite{merzbacher1958x} having values $c_K=c_{L_1}=1.5$;  $c_{L_2}=c_{L_3}=1.25$. The approximations of $I(x)$ is done as given in page (496) and equation (27) of \cite{brandt1979shell}, \cite{basbas1978universal} respectively.
The correction factor derived from the perturbed-stationary-state (PSS) theory with both binding and polarisation corrections capture cross-section formula is given in equation (20) of \cite{brandt1979shell}.The term $\zeta_s$ in this equation is written as $\epsilon_t$ in our paper. 

To incorporate the PSS correction factor in the capture cross-section formula we have to replace $\theta_t$ by $\epsilon_t\theta_t$ for the expression of minimum momentum transfer $q_t$ in equation (\ref{q(theta)}) and subsequently this $q_t$ is used in the the dimensionless parameter $\xi_t$ of equation (\ref{xi(theta)}).  

For M-shell electron capture, any binding and polarisation correction factor has not been developed in such a manner. To introduce the correction factor for Perturbed-stationary-state of M-shell electron Liu and Cipolla \cite{liu1996isics} have introduced united atom binding energy in equation (A.6) as $\zeta_s$ which we have written as $\lambda_t$ in our paper. So the modification for M-shell will be replacing $\theta_t$ by $\lambda_t\theta_t$ in $q_t$ and $\xi_t$ of equations (\ref{q(theta)}) and (\ref{xi(theta)}) respectively.

\subsection{Coulomb deflection correction}
The projectile ion has been described by a plane wave before and after interaction with Coulomb field of the target nucleous in case of Plane Wave Born Approximation (PWBA) which corresponds to a straight-line trajectory of the projectile ion in Semi-Classical-Approximation(SCA). However, this approximation becomes inadequate for slowly moving projectile ions (i.e. low velocity regime) because of the considerable distortion of the plane wave in Coulomb field of target nucleus. The experimental cross-sections for inner-shell vacancy creation in ion atom collision become significantly smaller than the prediction of the Plane-Wave-Born-Approximation. Bang and Hansteen have showed that for K-shell ionization, the theoretical prediction using Hyperbolic trajectory gives better agreement with experimental results.  
Lapicki and Losonsky \cite{lapicki1977electron} considered the Coulomb deflection correction factor as $e^{-\pi dq_t}$ where, $q_t$ and $d(=z_1z_2/\mu v^2)$ are the momentum transfer and half distance of closest approach respectively. Here, $\mu=\frac{M_pM_t}{M_p+M_t}$ is the reduced mass of target-projectile system. For electron capture process, the momentum transfer $q_t$ is different from that of the direct Coulomb ionisation to take into account the momentum of the transferred electron to the vacant shell of the projectile ion and the d is also modified to D (which is the symmentrized version of d) as prescribed by Lapicki et al.\cite{lapicki1977electron} and the $d_{ss'}$ of the paper is written as D here.

 The Coulomb deflection correction factor is hence expressed in equation (7) of \cite{lapicki1980electron}. Here, the d has been replaced by D and the equation has been written for K,L and M shells.
The minimum momentum transfer in the expression of Coulomb Deflection correction factor for K,L and M shells are $q_t(\epsilon_t\theta_t)$ and $q_t(\lambda_t\theta_t)$ that is $q_t$ induced by PSS.
The effect for the Coulomb deflection has been incorporated in the cross-section formula by simply multiplying the correction term $C_t$ to the cross-section. 
\subsection{Correction for Energy loss effect}
As it is clear from the direct Coulomb ionization theory, particularly for low projectile velocity regime, finite kinetic energy loss suffered by the projectile ions should be considered during the inner-shell ionization process to bridge the gap between the theory and experimental findings. We have incorporated the energy loss correction factor in capture cross-section formula as given by Brandt et al. in their theory  \cite{brandt1981energy}, \cite{liu1996isics}. The power of exponential in the coulomb Deflection correction factor is modified after considering the energy loss correction and is written in the bracketed term of equation (A.7) and the $Z_s$ is given in equation (A.8) of \cite{liu1996isics}. However, the $\xi_t(\theta_t)$ is replaced by $\xi_t(\epsilon_t\theta_t)$ (for K and L shells) or $\xi_t(\lambda_t\theta_t)$ (for M-shell) and instead of putting the simple ration $\xi/\zeta$.

\subsection{Relativistic correction}
The non-relativistic bound-state wave-function is simply expressed as the negative exponential of the distance from the nucleus i.e.  $exp(-r)$, however, this  form is modified to $r^{\gamma-1}exp(-r)$ where $\gamma^2=l-(z_2/137)^2$ for the relativistic case. Consequently, there are enhanced density of the high momentum components in the electronic momentum wave-function for  $r\rightarrow0$, which leads to the increase in ionization cross-section particularly for low velocity regime. So, the momentum transfer to the   target active shell electron can be calculated more efficiently with relativistic wave functions.\\
In the primary electron capture formula introduced by Oppenheimer-Brinkmann-Kramers with the modification of Nikolaev, i.e. in the OBKN approach (which is based on PWBA approximation), the non-relativistic electronic wave function has been used. 

Brandt and Lapicki \cite{brandt1979shell} have incorporated the relativistic correction in the theory of direct Coulomb ionization which we have included in the capture cross-section in a similar manner.
Here the, dimensionless factor $\xi_t$ is replaced with $\sqrt{m_t^R} \xi_t$, where $m_t^R$ is the relativistic correction factor as shown in \cite{brandt1979shell}, \cite{liu1996isics}. But in case of electron capture cross-section the correction factor can not be implemented by direct multiplication of the correction factor to $\xi_t$. For $K$ shell electron capture cross-section, Lapicki et al. have proposed a method to incorporate the relativistic correction, \cite{lapicki1980electron} where the cross-section formula is modified by replacing the projectile velocity $v$ with $v\sqrt{m_t^R}$. In this paper we are going to use this method for constructing a general relativistic cross-section formula for $K$ and $L$ shells.

Relativistic correction factor is considered in the OBKN approach which is essentially based on PWBA formalism (i.e. for high velocity regime) without considering the binding and polarisation factor, is known as $\sigma^{OBKNR}$, and written in equation (6) of \cite{brandt1979shell}. The term $m^R_s$ is written as $m^R_t$, the factor $\beta=1.1$ and the "c" has been replaced by 137 in atomic unit as prescribed by Lapicki et al. \cite{lapicki1980electron}.
The factor $\xi_{t}(\theta_t)$ is obtained from the equation (\ref{xi(theta)}).
\par In low velocity regime the perturbed stationary state approach is considered in the calculation of the relativistic correction factor $m^R_t(\xi_t(\epsilon_t\theta_t))$ which is obtained only by replacing $\xi(\theta_t)$ by $\xi(\epsilon_t\theta_t)$ in the forms of $m_t^R$ and $y_t$ and then the correction factor is included to the PSS induced momentum transfer in the similar manner of Lapicki's prescription.
\par
However, for target $M$-shell no calculation for relativistic correction is available, and $m_t^R$ is considered to be 1 in case of $M$-shell electron capture cross-section. 
\section{Electron capture cross-section for target K,L,M shell} 
After incorporating the relativistic factor without the perturbed stationary state approach we obtain the OBKN cross-section only with relativistic correction for high velocity limit of the projectile ion expressed as $\sigma^{OBKNR}_{K,L}$. For M-shell the relativistic correction factor $m_t^R(\xi_t(\theta_t))=1$ due to unavailability of calculation regarding M-shell resulting no change in OBKN cross-section. So, for M-shell the equation in high energy limit would be written as $\sigma^{OBKN}_{M}$.

For slow collision after incorporating all the correction factors the equation for K, L shell is given as $\sigma^{capture}_{K,L}(<)$. For M-shell the equation becomes $\sigma^{capture}_M(<)$

 A convenient formula has been proposed by Lapicki et al. to connect the low and high velocity limits of the projectile ions which can be expressed as in equation (10) of \cite{lapicki1980electron,lapicki1981erratum}
 
Now this formula is valid only for fully vacant projectile active shell where the electron is being transferred. Depending on the availability of vacancies in the projectile ion’s active shell, and the number of electrons in that shell in the ground state of the projectile atom a scaling factor $"f"$ has to be considered to calculate the ionization cross-section through electron capture mechanism. The correction term is incorporated in the cross-sections as a multiplicative factor. Using the above methodology, ionisation cross-section through target electron capture can be calculated not only for K-shell but also for sub-shell resolved L and M shells.    
 \section{Mean charge state and charge state distribution of the projectile ion inside target}
As we all know that after entering into the target material the projectile ions reaches an equilibrium charge state after passing a few layers of it through various phenomena like inner-shell ionization, several radiative and non-radiative processes which occurs during ion-atom collision. To compare the theoretical values with the experimental results, we need to know the mean charge state and the fractional charge state distribution of the projectile ions inside the target. 

Several analytical forms exist in the published literature for calculating the mean charge state of the projectile ions. However, these forms are applicable for the projectile ions after leaving the target material. Based on the experimental results the following two convenient models have been used to calculate the mean charge state of the projectile ions inside the target material.\\
\textbf{\underline{Fermi Gas Model:}}\\ We have used Fermi-Gas model \cite{brandt1973dynamic} to estimate the mean charge state of projectile ions inside the target especially for  in all projectile energy range for low z projectile and high projectile energy regime for high z-projectile ions and  $v_F<v$ which can be written as 
\begin{equation}q_m(FGM)\approx z_1(1-\frac{\gamma v_F}{v}).\label{q_f}\end{equation}
where, $v_F$ is the Fermi velocity of the target material, $v$ is the projectile velocity and $\gamma$ is a constant of the order of 1. \\
\textbf{\underline{Schiwietz Model:}}\\
An empirical formula for calculating the mean-charge-state of the projectile ion is proposed by Schiwietz et al. \cite{schiwietz2004femtosecond}
using least-square fitting to a large number of data points (800). In this model the target dependency of the projectile mean-charge-state comes from the two correction terms $c_1$ and $c_2$ as given in equations (3) and (4) of \cite{schiwietz2004femtosecond}.  The equation of the mean charge state obtained using Schiwietz Model is given as \begin{equation}q_m(Schiwietz)=z_1\frac{8.29x+x^4}{(0.06/x)+4+7.4x+x^4}, x=c_1(\frac{v_r}{1.54 c_2})^{1+1.83/z_1} \label{q_s}\end{equation}where, reduced velocity $v_r=z_1^{-0.543}\frac{v}{v_B}$, 
 $v$  and $v_B$ are projectile velocity and Bohr velocity respectively.
We have applied this formulation for high-z projectile and low projectile energy regime  for the available data sets that have been analysed in this paper. 

The  charge state distribution of the projectile ions follow Lorentzian nature inside the target material as  shown experimentally by Nandi et al. \cite{sharma2016experimental}.
The width of distribution curve has been calculated theoretically from Novikov and Teplova approach using equations (3) and (4) of \cite{novikov2014methods} 
where, $q_{av}$ and $Z=z_1$ are mean charge state and atomic number of projectile respectively.
In Fig. 1 we have shown the mean charge states of Ar ion inside Cu target for energy range 32-64 MeV along with the charge state distribution for 48 MeV energy.
\section{Simultaneous Multiple Ionisation (SMI) effect on Direct Coulomb Ionisation (DCI)}
During Direct Coulomb Ionization, Simultaneous Multiple ionization (SMI) of higher shells of the target atom due to the strong perturbation causes change in the probability of radiative and non-radiative transitions which essentially modify the atomic parameters like Fluorescence Yield and Coster-Kronig transition rate. To estimate the x-ray production cross-section from the theory of Direct Coulomb Ionization one should use the modified values of these atomic parameters for considering the effect of multiple ionization. 
 Here we have used the generalised formalism of the probability of SMI (P) as prescribed by Sulik \cite{sulik1987simple} on the basis of modified Binary Encounter Approximation (BEA) considering zero impact parameter. 
According to the prescription of Sulik \cite{sulik1987simple}  the expression of SMI probability is given as
\begin{equation}
P=P(X_n)= \frac{X_n^2}{4.2624+X_n^2[1+0.5~exp(-X_n^2/16)]}\label{SMI probability}
\end{equation}
Here, $X_n=W/n$ is the universal scaling parameter,  $n$ is the principle quantum number of the given target shell involved in SMI,
$W=4\frac{z_1}{v}V[G(V)]^{1/2}$ is  universal scaling variable, and $G(V)$ is function of scaled velocity $V=v/v_2$ \cite{mcguire1973procedure}.
The incorporation of SMI modify the atomic parameters as prescribed by Lapicki et al. \cite{lapicki1986multiple} as follows
\begin{equation}
\omega_i^{MI}
=\frac{\omega_i^0}{1-P(1-w_i^0)},\hspace{1cm} f_{ij}^{MI}=f_{ij}^0(1-P)^2\label{omegai,fij} 
\end{equation}
Where, $\omega_i^0$ and $f_{ij}^0$ are Fluorescence Yield and Coster-Kronig transition rate respectively for singly ionized atom.
\section{Comparison with experiments}
To verify our theoretical models for calculating Electron Capture cross-section we have used some experimental results obtained from well established literature of heavy ion induced inner-shell ionisation. 
We have followed the given sequence step by step to calculate the electron capture cross-sections. First, the mean charge state has been estimated using suitable empirical formulas as discussed in section-4. The width of the distribution curve has been determined as proposed by Novikov and Teplova approach and Lorentzian distribution has been considered to obtain the fractional charge-state distribution which provides the contribution of each charge state. We have calculated the electron capture cross-sections for different charge states for a given projectile energy and weighted those values by multiplying them with the fractional contributions of the corresponding charge states. Now we have to take the sum of the cross-sections of the different charge states for a particular energy and a particular sub-shell which essentially gives the theoretical data for sub-shell resolved  ionization cross-section through electron capture.

Theoretically calculated ionization cross-section through EC mechanism is converted to x-ray production cross-sections using several atomic parameters e.g. Fluorescence Yield $(\omega_i^o)$, Coster-Kronig transition rate $(f_{ij}^o)$, fractional radiative width which are  tabulated in various standard data tables for singly ionized atoms \cite{hubbell1994review,campbell2003fluorescence,chauhan2008mi,campbell1989interpolated,puri2007relative}, as described in  \cite{chatterjee2021significance, chatterjee2022understanding, mitra2010lower}  for different observed x-ray lines.

The modified values of Fluorescence Yield $(\omega_i^{MI})$, Coster-Kronig transition rates $(f_{ij}^{MI})$ as discussed in section-5 is used for considering the SMI effect for getting x-ray production cross-sections from the theoretically obtained ionization cross-sections through DCI process. The sum of the x-ray production cross-sections through EC mechanism and that through DCI along with the effect of Simultaneous Multiple Ionization has been compared with the experimental data obtained from various literature.

We also have calculated the total x-ray production cross-section from sub-shell resolved ionization cross-sections for both the ionizing channels (i.e., DCI and EC mechanism) as follows
\begin{equation}
\sigma^x_K=\omega_{K}\sigma_{K}^i\label{K-total production}
\end{equation}
\begin{equation}
\sigma^x_{L_{total}}=\omega_{L_1}\sigma_{L_1}^i+\omega_{L_2}(\sigma_{L_2}^i+f_{12}\sigma_{L_1}^i)+\omega_{L_3}[\sigma_{L_3}+f_{23}\sigma_{L_2}^i+(f_{13}+f_{12}f_{23})\sigma_{L_1}^i] \label{L-total production}
\end{equation}  
\begin{equation}
    \begin{split}
\sigma^x_{M_{total}}=\omega_{M_1}\sigma_{M_1}^i+\omega_{M_2}(\sigma_{M_2}^i+f_{12}\sigma_{M_1}^i)+\omega_{M_3}[\sigma_{M_3}^i+f_{23}\sigma_{M_2}^i+(f_{13}+f_{12}f_{23})\sigma_{M_1}^i]+&\\\omega_{M_4}[\sigma_{M_4}^i+f_{34}\sigma_{M_3}^i+(f_{24}+f_{23}f_{34})\sigma_{M_2}^i+(f_{14}+f_{13}f_{34}+f_{12}f_{24}+f_{12}f_{23}f_{34})\sigma_{M_1}^i]+&\\\omega_{M_5}[\sigma_{M_5}^i+\sigma_{M_4}^if_{45}+\sigma_{M_3}^i(f_{35}+f_{34}f_{45})+\sigma_{M_2}^i(f_{25}+f_{23}f_{35}+f_{24}f_{45}+f_{23}f_{34}f_{45})+&\\\sigma_{M_1}^i(f_{15}+ f_{12} f_{25}+ f_{13} f_{35}+ f_{14} f_{45}+ f_{12}f_{23}f_{35}+ f_{12} f_{24} f_{45}+ f_{13} f_{34} f_{45}+ f_{12} f_{23} f_{34} f_{45})]\label{M-total production}
    \end{split}
\end{equation}
\par
{Here, we have considered K-shell ionization cross-sections for Si induced Ti \cite{msimanga2016k} and Ar induced Cu and Zn \cite{gluchshenko2016k} target and the comparison has been graphically shown in Fig. 2. As $z_1>10$ (projectile atomic number) and projectile energy $E_p\ge0.4MeV/u$, we have used Fermi-Gas Model to determine the mean charge state of the projectile ions with Fermi Velocity ($v_F $) 2.18, 2.6 and 2.75 in atomic unit for Ti, Cu and Zn targets respectively. After incorporating the Electron Capture mechanism along with the Simultaneous Multiple Ionization effect in Direct Coulomb Ionization, though the theoretical prediction of x-ray production cross-sections of Copper target slightly over predict the experimental data for high energy regime, overall improvement is quite satisfactory for all target-projectile combinations. 

Similarly, the comparative studies between the theory and experiment have been shown in figures (Fig.3, Fig.4, Fig.5) in case of L-shell ionization. Here, sub-shell resolved L-shell ionization for Bi, Ta and Pb target induced by Ar projectile ion have been taken into consideration from literature \cite{gluchshenko2016k} as experimental data. It is clear from the figures that the overall theoretically estimated total x-ray production cross-sections show better agreement with the experimental findings after considering the ionization through EC mechanism along with the SMI effect in DCI. However, in the sub-shell resolved cases few discrepancies are noted which may be due to the uncertainties in used atomic parameters for singly ionized atom as well as the from estimation of the extent of multiple ionization. 

Unlike K and L shell, M-shell has five sub-shells with different properties. To verify our theoretical model more precisely we have compared the experimental results in case of M shell. X-ray production cross-section for few heavy-z targets induced by C, Si, S and Ar projectile ions have been compared with the theoretical predictions \cite{mitra2010lower,mitra2001m,wang2012multiple} and shown in Fig. 6, 7, 8 and 9. Mean charge state of the C projectile ions is determined using Fermi-Gas Model as discussed in Section 4. It can clearly be observed from Fig.6, though the theoretical prediction improves significantly when compared to experimental data, it under-predicts in case of Pb target. For Si, S and Ar projectile ion, and low energy range the mean charge state is estimated using Schiwietz Model. It is clear from Fig.7, Fig.8 and Fig.9 that though the theoretical prediction underestimates the experimental data for few target projectile combinations, overall significant improvement is observed after incorporation of the contribution of ionization through electron capture mechanism. The mismatch between the experimental data and theoretical predictions may be due to the inaccuracy in the theoretically estimated different atomic parameters (e.g. $\omega_i, f_{ij}$, fractional radiative width), which are used to convert the ionization cross-sections to x-ray production cross-sections. Accurate knowledge of the probability of Simultaneous Multiple Ionization and their effect on atomic parameters for the case of Direct Coulomb ionization is also required for the calculation of x-ray production cross-sections theoretically. Another cause of discrepancies between the theory and experiment maybe due to the anisotropic emission of the M x-rays \cite{mitra1998measurement} through heavy ion induced inner-shell ionization. All the measurements for M x-ray production cross-sections had been performed by putting the x-ray detectors at $90^0$ with the beam direction. So the knowledge of the anisotropic parameter and its energy dependence is required for correcting the experimentally measured x-ray production cross-sections before comparing with the theoretical predictions. 

To understand the discrepancies between the theory and experiments in more details, further experimental investigations are required in this direction for target K, L and M shells.} 

 \section*{Acknowledgement}
 One of the authors, SG, acknowledges the University
of Kalyani for providing her the SVMCM scholarship during this research work.
\bibliography{main.bib}

\begin{thebibliography}{10}

\bibitem{macdonald1972dependence}
James~R Macdonald, Loren Winters, Matt~D Brown, Tang Chiao, and Louis~D Ellsworth.
\newblock {Dependence of X-Ray Yields in Argon, Krypton, and Xenon upon the Charge State of Fluorine Ions at 35.7 MeV}.
\newblock {\em Physical Review Letters}, 29(19):1291, 1972.

\bibitem{mowat1972projectile}
J~Richard Mowat, DJ~Pegg, RS~Peterson, PM~Griffin, and IA~Sellin.
\newblock {Projectile Structure Effects on Neon K X-ray production by fast, highly ionized Argon beams}.
\newblock {\em Physical Review Letters}, 29(24):1577, 1972.

\bibitem{brandt1973dynamic}
Werner Brandt, Roman Laubert, Manuel Mourino, and Arthur Schwarzschild.
\newblock Dynamic screening of projectile charges in solids measured by target x-ray emission.
\newblock {\em Physical Review Letters}, 30(9):358, 1973.

\bibitem{basbas1971projectile}
George Basbas, Werner Brandt, Roman Laubert, Anthony Ratkowski, and Arthur Schwarzschild.
\newblock Projectile charge dependence of {K}-shell ionization by swift light nuclei.
\newblock {\em Physical Review Letters}, 27(4):171, 1971.

\bibitem{brandt1966binding}
W~Brandt, R~Laubert, and I~Sellin.
\newblock Binding effects in electronic excitations by heavy charged particles.
\newblock {\em Physics Letters}, 21(5):518--519, 1966.

\bibitem{brandt1966characteristic}
Werner Brandt, Roman Laubert, and Ivan Sellin.
\newblock {Characteristic x-ray production in Magnesium, Aluminum, and Copper by low-energy Hydrogen and Helium ions}.
\newblock {\em Physical Review}, 151(1):56, 1966.

\bibitem{msimanga2016k}
M~Msimanga, CA~Pineda-Vargas, and M~Madhuku.
\newblock {K-shell X-ray production cross sections in Ti by 0.3-1.0 MeV/u $^{12}$C and $^{28}$Si ions for heavy ion PIXE}.
\newblock {\em Nuclear Instruments and Methods in Physics Research Section B: Beam Interactions with Materials and Atoms}, 380:90--93, 2016.

\bibitem{gluchshenko2016k}
N~Gluchshenko, I~Gorlachev, I~Ivanov, A~Kireyev, S~Kozin, A~Kurakhmedov, A~Platov, and M~Zdorovets.
\newblock {K-, L-and M-shell X-ray productions induced by argon ions in the 0.8-1.6 MeV/amu range}.
\newblock {\em Nuclear Instruments and Methods in Physics Research Section B: Beam Interactions with Materials and Atoms}, 372:1--6, 2016.

\bibitem{mitra2010lower}
D~Mitra, M~Sarkar, D~Bhattacharya, S~Santra, AC~Mandal, and G~Lapicki.
\newblock {Lower and upper bounds on M-shell X-ray production cross sections by heavy ions}.
\newblock {\em Nuclear Instruments and Methods in Physics Research Section B: Beam Interactions with Materials and Atoms}, 268(5):450--459, 2010.

\bibitem{richard1969observed}
Patrick Richard, IL~Morgan, T~Furuta, and D~Burch.
\newblock {Observed K $\beta$ Energy Shift in Cu and Ni}.
\newblock {\em Physical Review Letters}, 23(18):1009, 1969.

\bibitem{burch1970x}
D~Burch and Patrick Richard.
\newblock {X-ray spectra from Oxygen-ion bombardments on Ca and V at 15 MeV}.
\newblock {\em Physical Review Letters}, 25(15):983, 1970.

\bibitem{knudson1971aluminum}
AR~Knudson, DJ~Nagel, PG~Burkhalter, and KL~Dunning.
\newblock Aluminum x-ray satellite enhancement by ion-impact excitation.
\newblock {\em Physical Review Letters}, 26(19):1149, 1971.

\bibitem{richard1972direct}
Patrick Richard, W~Hodge, and C~Fred Moore.
\newblock {Direct Observation of K $\alpha$ Hypersatellites in Heavy-Ion Collisions}.
\newblock {\em Physical Review Letters}, 29(7):393, 1972.

\bibitem{burch1972effect}
D~Burch, WB~Ingalls, JS~Risley, and R~Heffner.
\newblock {Effect of Multiple Ionization on the Fluorescence Yield of Ne}.
\newblock {\em Physical Review Letters}, 29(26):1719, 1972.

\bibitem{halpern1973k}
AM~Halpern and J~Law.
\newblock {K-vacancy creation by high-Z heavy-ion impact}.
\newblock {\em Physical Review Letters}, 31(1):4, 1973.

\bibitem{nikolaev1967calculation}
VS~Nikolaev.
\newblock Calculation of the effective cross sections for proton charge exchange in collisions with multi-electron atoms.
\newblock {\em Sov. Phys. JETP}, 24(847):163, 1967.

\bibitem{lapicki1977electron}
Grzegorz Lapicki and William Losonsky.
\newblock Electron capture from inner shells by fully stripped ions.
\newblock {\em Physical Review A}, 15(3):896, 1977.

\bibitem{lapicki1980electron}
Gregory Lapicki and Floyd~Del McDaniel.
\newblock Electron capture from {K} shells by fully stripped ions.
\newblock {\em Physical Review A}, 22(5):1896, 1980.

\bibitem{brandt1981energy}
Werner Brandt and Grzegorz Lapicki.
\newblock Energy-loss effect in inner-shell {C}oulomb ionization by heavy charged particles.
\newblock {\em Physical Review A}, 23(4):1717, 1981.

\bibitem{slater1930atomic}
John~C Slater.
\newblock Atomic shielding constants.
\newblock {\em Physical Review}, 36(1):57, 1930.

\bibitem{agmon1988ionization}
Noam Agmon.
\newblock Ionization potentials for isoelectronic series.
\newblock {\em Journal of Chemical Education}, 65(1):42, 1988.

\bibitem{liu1996isics}
Zhiqiang Liu and Sam~J Cipolla.
\newblock {ISICS: A program for calculating K-, L-and M-shell cross sections from ECPSSR theory using a personal computer}.
\newblock {\em Computer Physics Communications}, 97(3):315--330, 1996.

\bibitem{basbas1973perturbed}
George Basbas, Werner Brandt, and RH~Ritchie.
\newblock Perturbed-stationary-state theory of atomic inner-shell ionization by heavy charged particles.
\newblock {\em Physical Review A}, 7(6):1971, 1973.

\bibitem{brandt1979shell}
Werner Brandt and Grzegorz Lapicki.
\newblock L-shell {C}oulomb ionization by heavy charged particles.
\newblock {\em Physical Review A}, 20(2):465, 1979.

\bibitem{merzbacher1958x}
E~Merzbacher and HW~Lewis.
\newblock X-ray production by heavy charged particles.
\newblock In {\em Corpuscles and Radiation in Matter II/Korpuskeln und Strahlung in Materie II}, pages 166--192. Springer, 1958.

\bibitem{basbas1978universal}
George Basbas, Werner Brandt, and Roman Laubert.
\newblock {Universal cross sections for K-shell ionization by heavy charged particles. II. Intermediate particle velocities}.
\newblock {\em Physical Review A}, 17(5):1655, 1978.

\bibitem{lapicki1981erratum}
G~Lapicki and FD~McDaniel.
\newblock Erratum: Electron capture from k shells by fully stripped ions.
\newblock {\em Physical Review A}, 23(2):975, 1981.

\bibitem{schiwietz2004femtosecond}
G~Schiwietz, K~Czerski, M~Roth, F~Staufenbiel, and PL~Grande.
\newblock Femtosecond dynamics--snapshots of the early ion-track evolution.
\newblock {\em Nuclear Instruments and Methods in Physics Research Section B: Beam Interactions with Materials and Atoms}, 225(1-2):4--26, 2004.

\bibitem{sharma2016experimental}
Prashant Sharma and Tapan Nandi.
\newblock Experimental evidence of beam-foil plasma creation during ion-solid interaction.
\newblock {\em Physics of Plasmas}, 23(8):083102, 2016.

\bibitem{novikov2014methods}
NV~Novikov and Ya~A Teplova.
\newblock Methods of estimation of equilibrium charge distribution of ions in solid and gaseous media.
\newblock {\em Physics Letters A}, 378(18-19):1286--1289, 2014.

\bibitem{sulik1987simple}
B~Sulik, I~K{\'a}d{\'a}r, S~Ricz, D~Varga, J~V{\'e}gh, G~Hock, and Denes Ber{\'e}nyi.
\newblock {A simple theoretical approach to multiple ionization and its application for 5.1 and 5.5 MeV/u $X^{q+}-$ Ne collisions}.
\newblock {\em Nuclear Instruments and Methods in Physics Research Section B: Beam Interactions with Materials and Atoms}, 28(4):509--518, 1987.

\bibitem{mcguire1973procedure}
James~H McGuire and Patrick Richard.
\newblock Procedure for computing cross sections for single and multiple ionization of atoms in the binary-encounter approximation by the impact of heavy charged particles.
\newblock {\em Physical Review A}, 8(3):1374, 1973.

\bibitem{lapicki1986multiple}
Gregory Lapicki, R~Mehta, Jerome~L Duggan, PM~Kocur, JL~Price, and Floyd~Del McDaniel.
\newblock Multiple outer-shell ionization effect in inner-shell x-ray production by light ions.
\newblock {\em Physical Review A}, 34(5):3813, 1986.

\bibitem{hubbell1994review}
JH~Hubbell, PN~Trehan, Nirmal Singh, B~Chand, D~Mehta, ML~Garg, RR~Garg, Surinder Singh, and S~Puri.
\newblock {A review, bibliography, and tabulation of K, L, and higher atomic shell x-ray fluorescence yields}.
\newblock {\em Journal of Physical and Chemical Reference Data}, 23(2):339--364, 1994.

\bibitem{campbell2003fluorescence}
JL~Campbell.
\newblock {Fluorescence yields and Coster-Kronig probabilities for the atomic L subshells}.
\newblock {\em Atomic Data and Nuclear Data Tables}, 85(2):291--315, 2003.

\bibitem{chauhan2008mi}
Yogeshwar Chauhan and Sanjiv Puri.
\newblock {$M_i (i= 1-5)$ subshell fluorescence and Coster-Kronig yields for elements with 67$\le$ Z $\le$ 92}.
\newblock {\em Atomic Data and Nuclear Data Tables}, 94(1):38--49, 2008.

\bibitem{campbell1989interpolated}
JL~Campbell and J-X Wang.
\newblock {Interpolated Dirac-Fock values of L-subshell x-ray emission rates including overlap and exchange effects}.
\newblock {\em Atomic data and nuclear data tables}, 43(2):281--291, 1989.

\bibitem{puri2007relative}
Sanjiv Puri.
\newblock {Relative intensities for $L_i$ (i= 1-3) and $M_i$ (i= 1-5) subshell X-rays}.
\newblock {\em Atomic Data and Nuclear Data Tables}, 93(5):730--741, 2007.

\bibitem{chatterjee2021significance}
Soumya Chatterjee, Prashant Sharma, Shashank Singh, Mumtaz Oswal, Sunil Kumar, CC~Montanari, D~Mitra, and T~Nandi.
\newblock Significance of the high charge state of projectile ions inside the target and its role in electron capture leading to target-ionization phenomena.
\newblock {\em Physical Review A}, 104(2):022810, 2021.

\bibitem{chatterjee2022understanding}
Soumya Chatterjee, Sunil Kumar, Sarvesh Kumar, M~Oswal, Biraja Mohanty, D~Mehta, D~Mitra, AMP Mendez, Dario~Marcelo Mitnik, Claudia~Carmen Montanari, et~al.
\newblock Understanding the mechanisms of l-shell x-ray emission from os atoms bombarded by 4--6 mev/u fluorine ion.
\newblock {\em Physica Scripta}, 97(4):045405, 2022.

\bibitem{mitra2001m}
D~Mitra, AC~Mandal, M~Sarkar, D~Bhattacharya, P~Sen, and G~Lapicki.
\newblock M {X}-ray production cross-sections of gold and lead by 4 to 12 {MeV} carbon ions.
\newblock {\em Nuclear Instruments and Methods in Physics Research Section B: Beam Interactions with Materials and Atoms}, 183(3-4):171--177, 2001.

\bibitem{wang2012multiple}
Xing Wang, Yongtao Zhao, Rui Cheng, Xianming Zhou, Ge~Xu, Yuanbo Sun, Yu~Lei, Yuyu Wang, Jieru Ren, Yang Yu, et~al.
\newblock {Multiple ionization effects in M X-ray emission induced by heavy ions}.
\newblock {\em Physics Letters A}, 376(14):1197--1200, 2012.

\bibitem{mitra1998measurement}
D~Mitra, M~Sarkar, D~Bhattacharya, P~Sen, and G~Kuri.
\newblock Measurement of the anisotropy parameters for the m x-rays of gold induced by 3--9 mev carbon ions.
\newblock {\em Nuclear Instruments and Methods in Physics Research Section B: Beam Interactions with Materials and Atoms}, 145(3):283--287, 1998.

\end{thebibliography}
\bibliographystyle{unsrt}
\clearpage
\begin{figure}
\centering
\includegraphics[width=23cm,height=13cm]{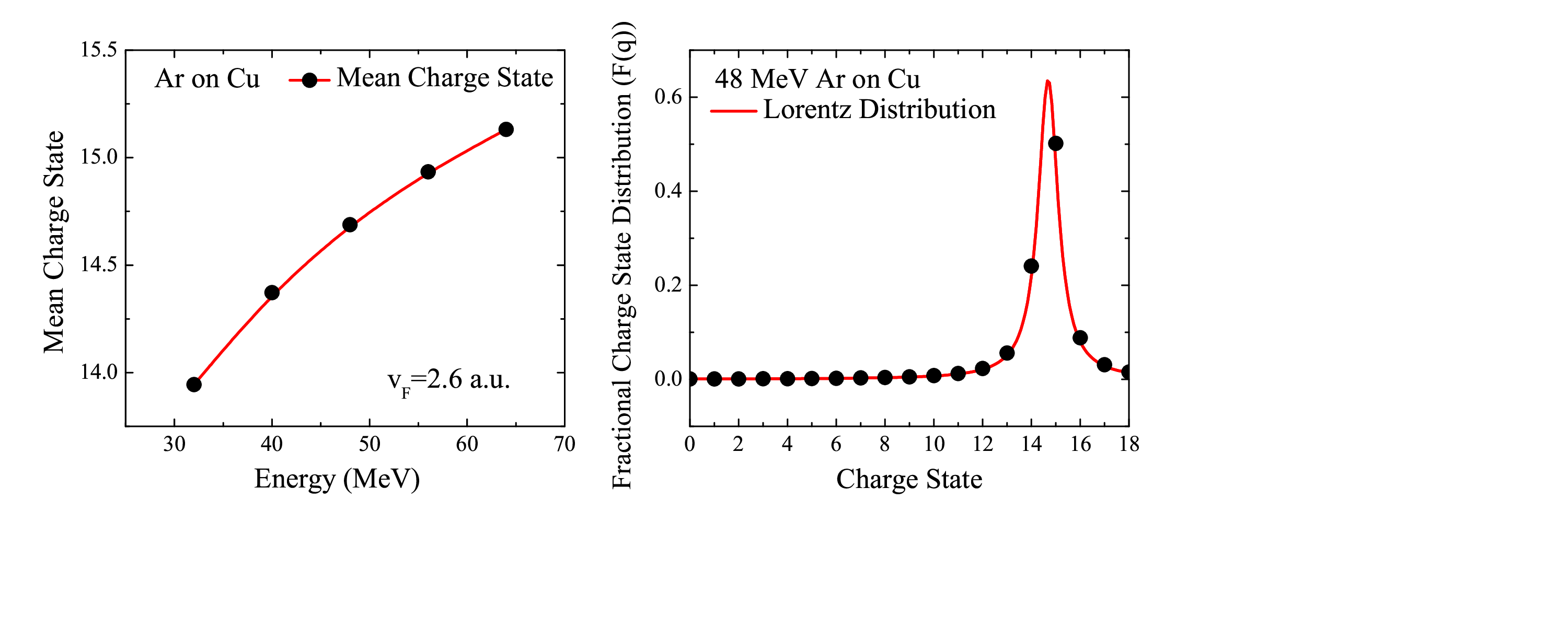}
\caption{Mean Charge State of Ar ion inside Cu target for energy range from 32 to 64 MeV(LHS) and Charge state distribution of 48 MeV Ar ions inside Cu target(RHS).}
\label{mean charge state and charge state distribution}.
\end{figure}
\begin{figure}
\centering
\includegraphics[width=15cm,height=20cm]{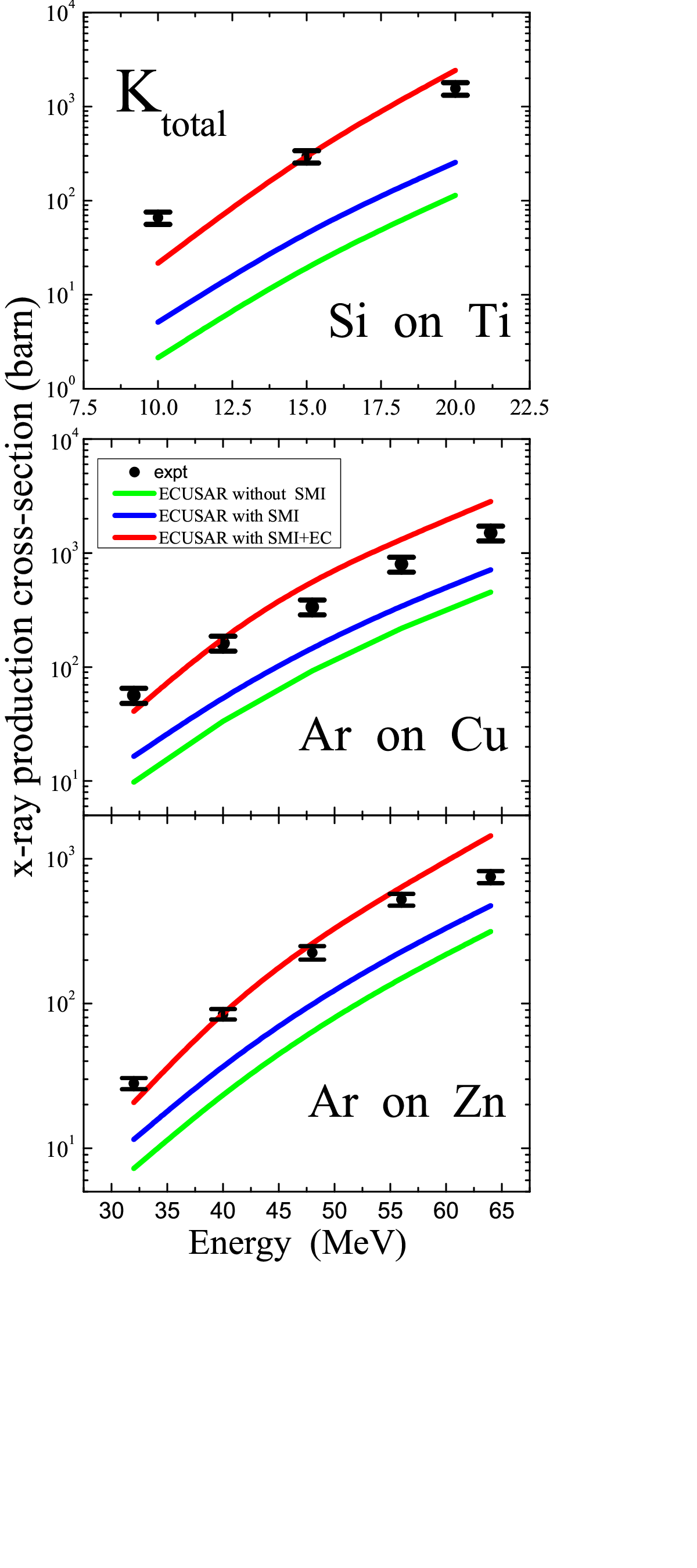}
\caption{K-shell x-ray production cross-section of Ti target induced by 10-20 MeV Si ions and Cu and Zn target induced by 32-64 MeV Ar ions. The experimental data is represented by solid dots where as the solids lines are obtained from different theories.}
\label{SIGMA-K}.
\end{figure}
\begin{figure}
\centering
\includegraphics[width=23cm,height=20cm]{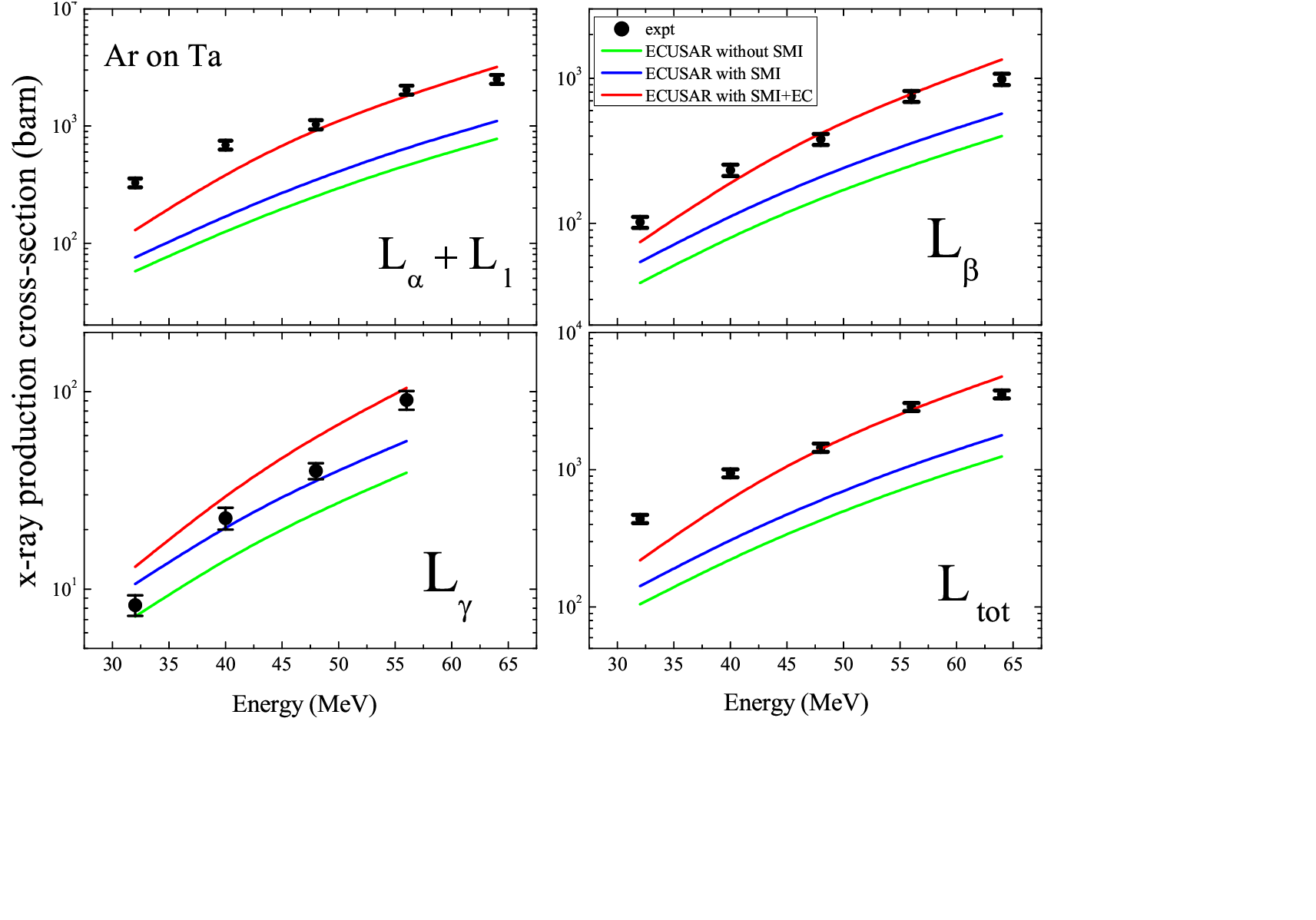}
\caption{L-shell x-ray production cross-section $(L_{\alpha+l}, L_{\beta}, L_{\gamma}$ and $ L_{total})$ for Ta target  induced by 32-64 MeV Ar projectile ions. The solid dots represent  the experimental findings and different solid lines corresponds to the data obtained from different theories.}.
\label{SIGMA-L-Ar-Ta}
\end{figure}
\begin{figure}
\centering
\includegraphics[width=23cm,height=20cm]{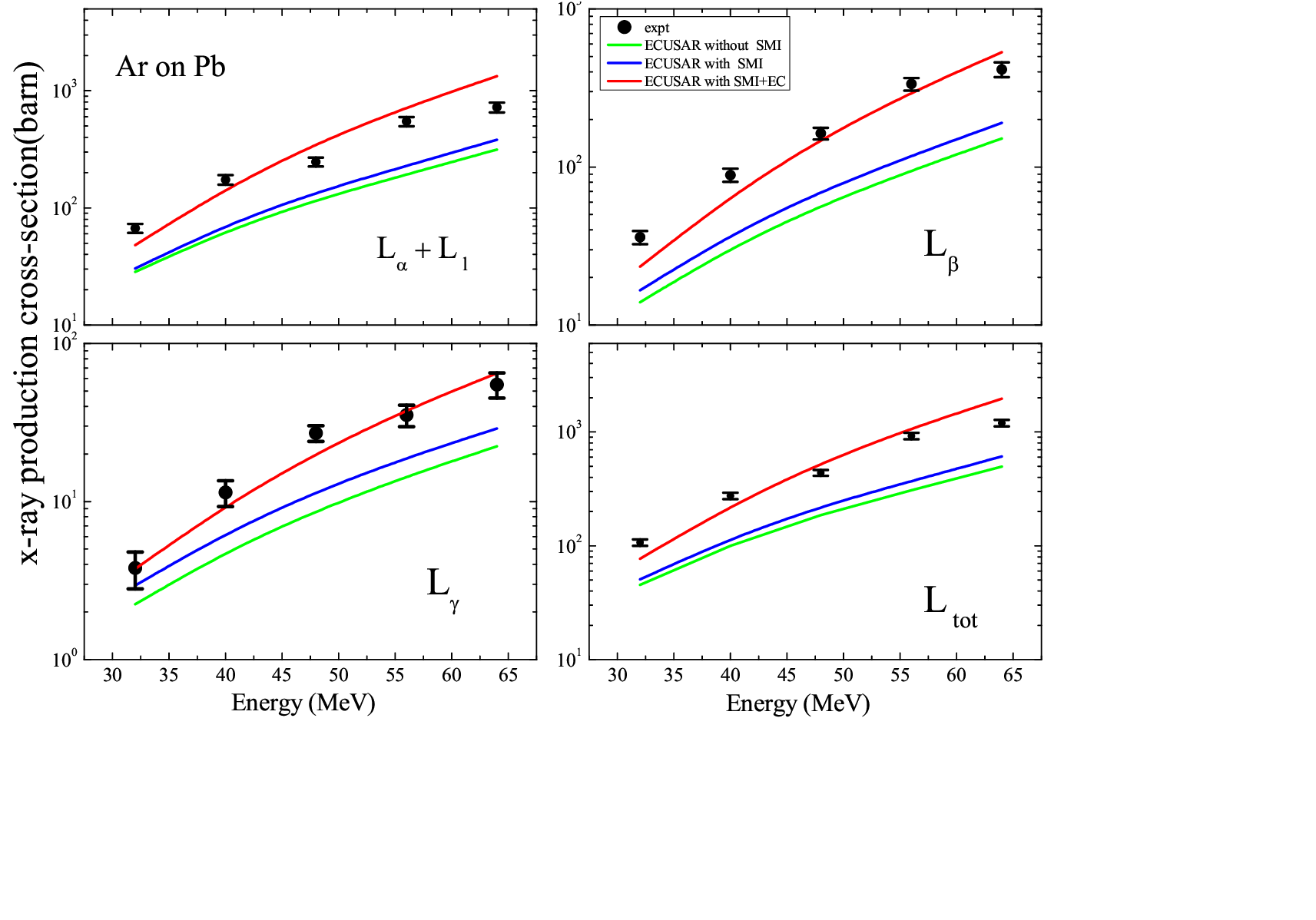}
\caption{L-shell x-ray production cross-section $(L_{\alpha+l}, L_{\beta}, L_{\gamma}$ and $ L_{total})$ for by 32-64 MeV Ar projectile ions induced Pb target. The solid dots represent  the experimental data and the solid lines corresponds to the values obtained from different theories.}.
\label{SIGMA-L-Ar-Pb}
\end{figure}
\begin{figure}
\centering
\includegraphics[width=23cm,height=20cm]{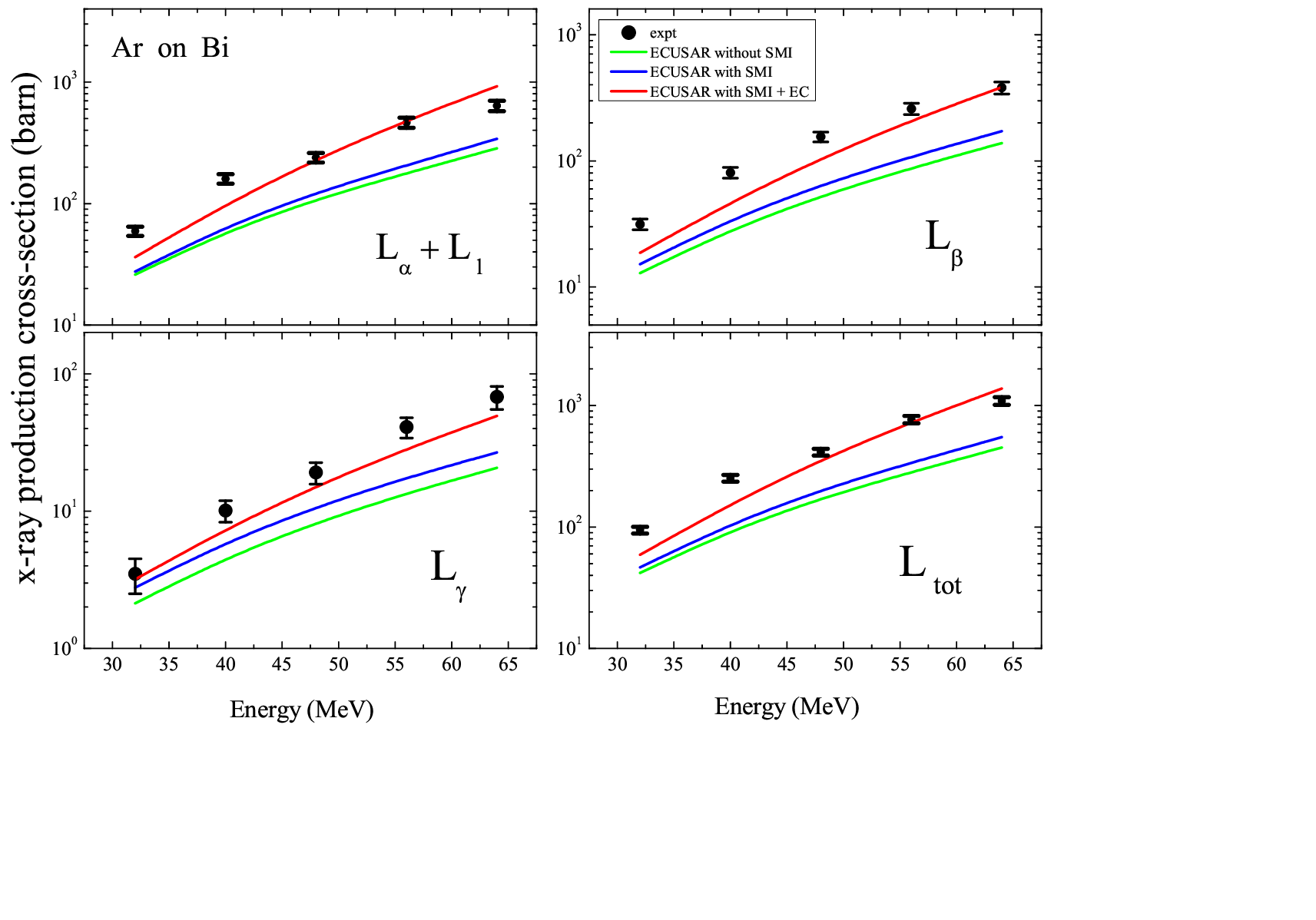}
\caption{L-shell x-ray production cross-section $(L_{\alpha+l}, L_{\beta}, L_{\gamma}$ and $ L_{total})$ for Bi target  induced by 32-64 MeV Ar projectile ions. The experimental findings are represented by the solid dots and the theoretical cross-section of various theories are specified by the solid lines.}.
\label{SIGMA-L-Ar-Bi}
\end{figure}
\begin{figure}
\centering
\includegraphics[width=23cm,height=23cm]{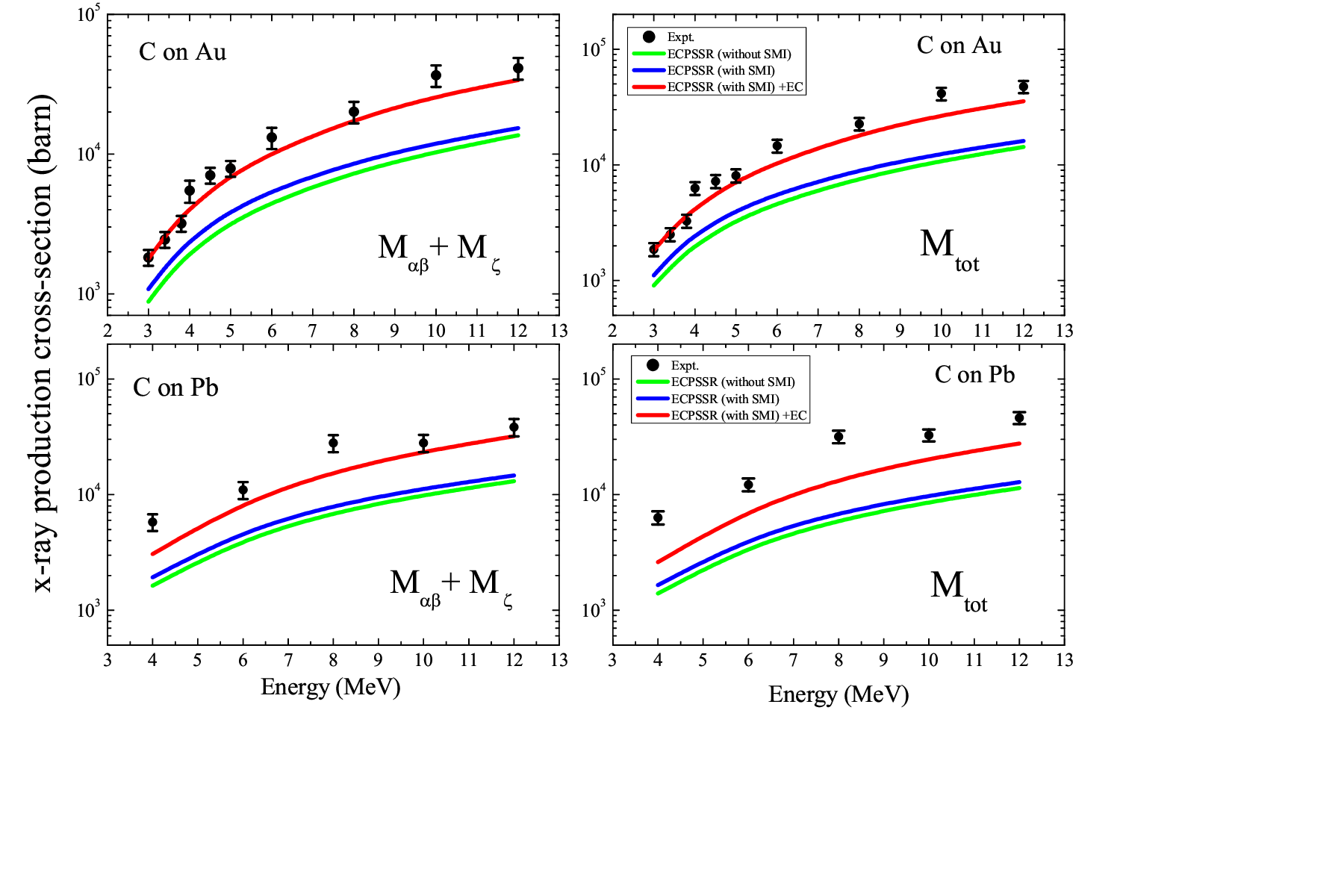}
\caption{ M-shell x-ray production lines $(M_{\alpha\beta+\zeta}$ and $M_{total})$ of Au and Pb target induced by C projectile ions. Solid dots and different solid lines, respectively, indicate the the experiment data and values determined theoretically.}.
\label{SIGMA-M-C-Au-Pb}
\end{figure} 
\begin{figure}
\centering
\includegraphics[width=23cm,height=23cm]{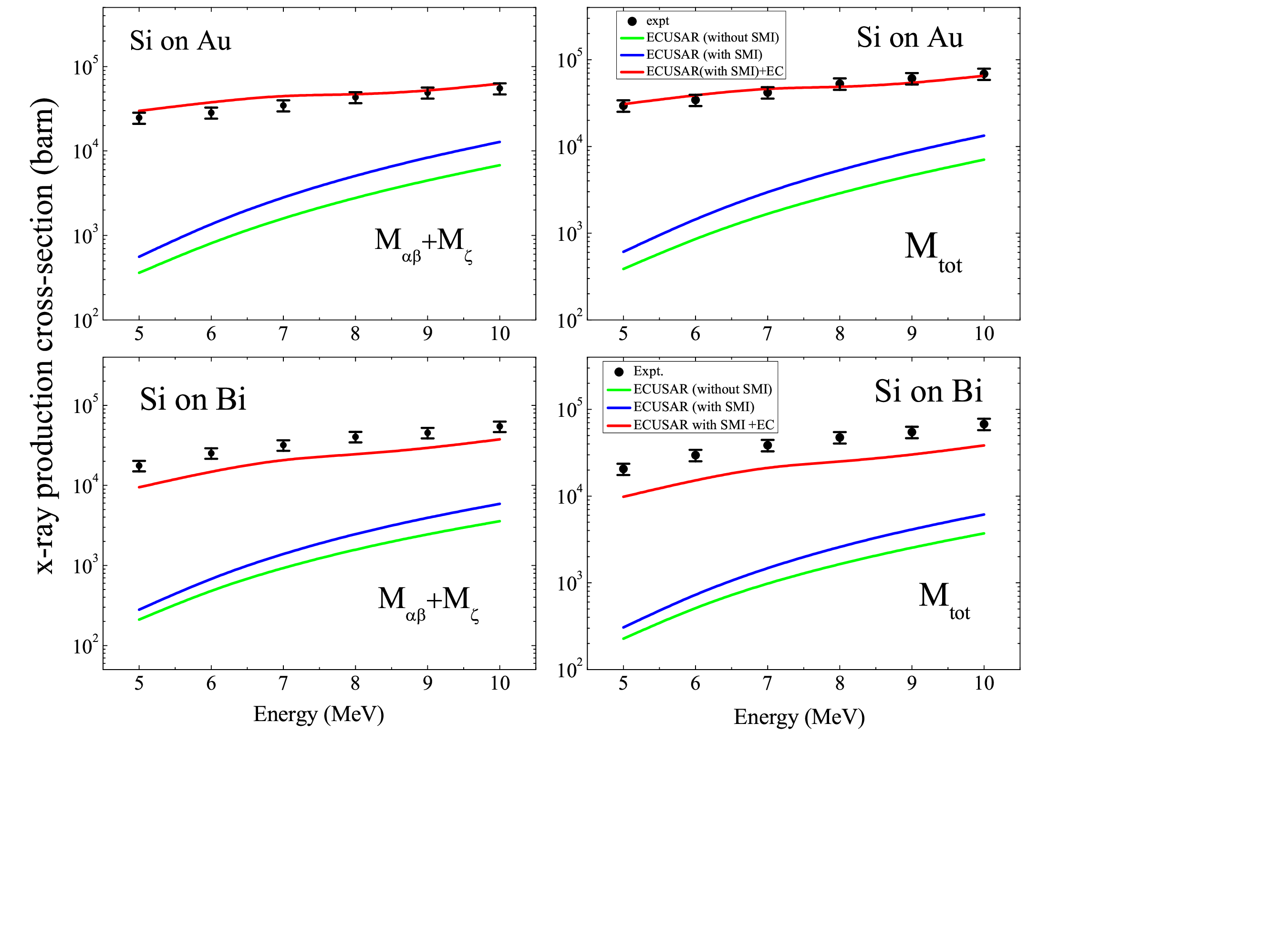}
\caption{ Si projectile ion induced Au and Bi targets M-shell x-ray production cross-section $(M_{\alpha\beta+\zeta}$ and $M_{total})$. The theoretically determined values are represented by different solid lines whereas the solid dots indicate the experimental data.}.
\label{Si on Au M}
\end{figure}
\begin{figure}
\centering
\includegraphics[width=23cm,height=23cm]{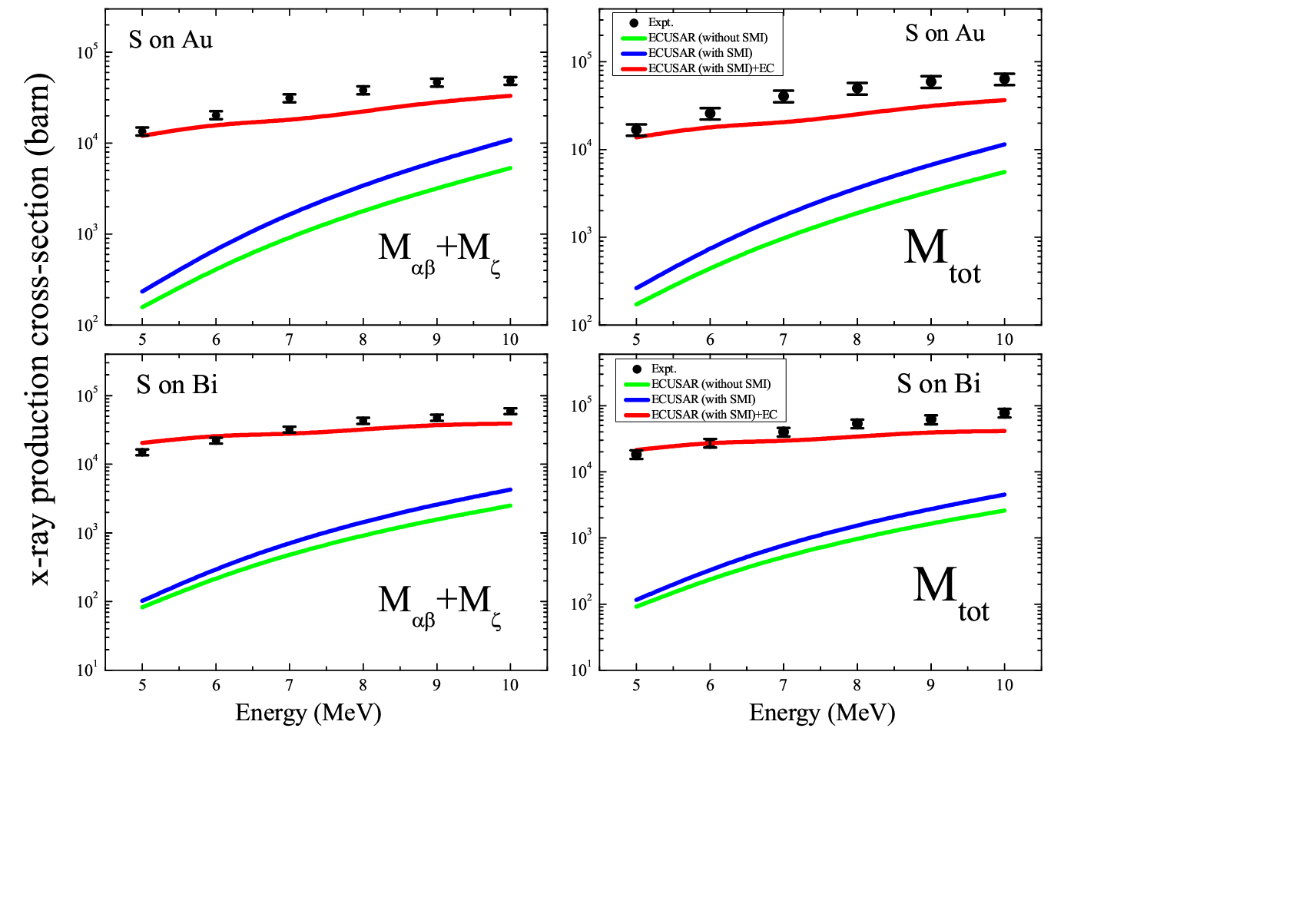}
\caption{S projectile ion induced Au and Bi targets M-shell x-ray production cross-section $(M_{\alpha\beta+\zeta}$ and $M_{total})$. The theoretical data are represented by different solid lines and the solid dots indicate the experimental data. }.
\label{S on Au M}
\end{figure}
\begin{figure}
\centering
\includegraphics[width=23cm,height=13cm]{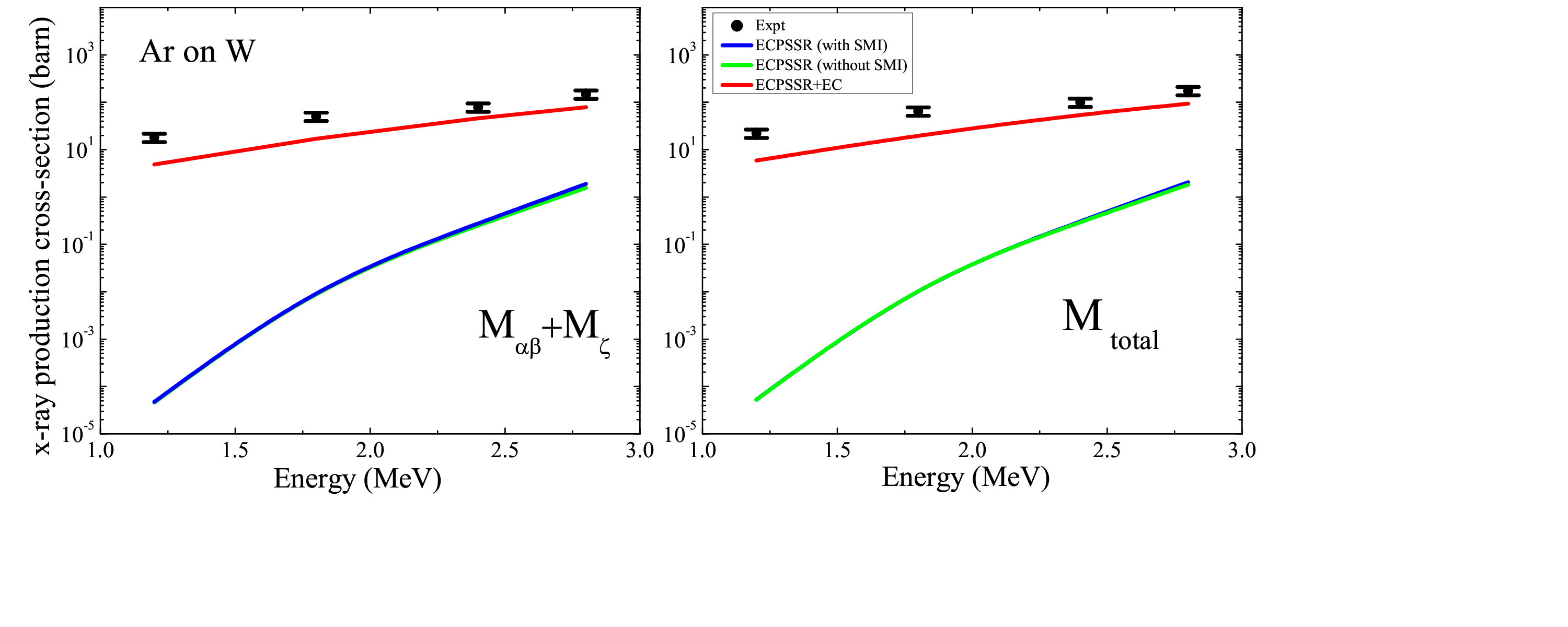}
\caption{M-shell x-ray production cross-section $(M_{\alpha\beta+\zeta}$ and $M_{total})$ of Ar projectile ion induced W target. The solid lines represent the theoretical values whereas the experimental data is specified by the solid dots. }.
\label{SIGMA-M-Ar-W}
\end{figure}
\end{document}